\documentclass[12pt]{article}

\usepackage{newtxtext,newtxmath}

\usepackage{graphicx}

\usepackage[letterpaper,margin=1in]{geometry}

\renewenvironment{abstract}
	{\quotation}
	{\endquotation}

\date{}

\makeatletter
\renewcommand{\fnum@figure}{\textbf{Figure \thefigure}}
\renewcommand{\fnum@table}{\textbf{Table \thetable}}
\makeatother

\usepackage{scicite}

\usepackage{url}

\usepackage[T1]{fontenc}

\def\scititle{
	     First-in-human quantum entanglement imaging 
}
\title{\bfseries \boldmath \scititle}

\author{
    Pawel~Moskal$^{1,2,3\ast}$,
	Deepak~Kumar$^{1,2,3}$,
    Sushil~Sharma$^{1,2,3}$,
    Ermias Y. Beyene$^{1,2,3}$,\and
    Neha~Chug$^{1,2,3}$, 
    Catalina~Curceanu$^{4}$,
    Eryk~Czerwi{\'n}ski$^{1,2,3}$,
    Atharva~Dalvi$^{1,2,3}$,\and
    Manish~Das$^{1,2,3}$,
    Alicja Hubalewska-Dydejczyk$^{5}$,
    Sharareh Jalali$^{1,2,3}$,\and
    Krzysztof~Kacprzak$^{1,2,3}$,
    Tevfik~Kaplanoglu$^{1,2,3}$,
    {\L}ukasz~Kap{\l}on$^{1,2,3}$,\and
    Kamila Kasperska$^{1,2,3}$,
    Aleksander Khreptak$^{1,2,3}$,
    Grzegorz Korcyl$^{1,2,3}$,\and
    Tomasz~Kozik$^{1,2,3}$,
    Sumit Kumar Kundu$^{1,2,3}$,
    Anoop Kunimmal Venadan$^{1,2,3}$,\and
    Bartosz Leszczyński$^{1,2,3}$,
    Edward~Lisowski$^{6}$,
    Filip~Lisowski$^{6}$,\and
    Justyna Mędrala-Sowa$^{1,2,3}$,
    Simbarashe Moyo$^{1,2,3}$,
    Wiktor~Mryka$^{1,2,3}$,\and
    Szymon~Nied{\'z}wiecki$^{1,2,3}$,
    Marta Opalińska$^{5}$,
    Anand Pandey$^{1,2,3}$,\and
    Piyush Pandey$^{1,2,3}$,
    Alessio Porcelli$^{1,2,3}$,
    Bartłomiej Rachwał$^{7}$,\and
    Magdalena~Skurzok$^{1,2,3}$,
    Anna Sowa-Staszczak$^{5}$,
    Tomasz Szumlak$^{7}$,\and
    Satyam Tiwari$^{1,2,3}$,
    Pooja~Tanty$^{1,2,3}$,
    Keyvan~Tayefi~Ardebili$^{1,2,3}$,\and
    Kavya~Valsan~Eliyan$^{1,2,3}$,
    Ewa~{\L{}}.~Stepie{\'n}$^{1,2,3}$\and
	\small$^{1}$Faculty of Physics, Astronomy and Applied Computer Science, Jagiellonian University, \and \small S.~Łojasiewicza 11, 30-348 Kraków, Poland.\and
	\small$^{2}$Total-Body Jagiellonian-PET Laboratory, Jagiellonian University, Poland.\and
    \small$^{3}$Center for Theranostics, Jagiellonian University, 31-034 Kraków, Poland.\and
    \small$^{4}$INFN, Laboratori Nazionali di Frascati CP 13,  Via E. Fermi 40, 00044, Frascati, Italy.\and
    \small$^{5}$Chair and Department of Endocrinology, Jagiellonian University Medical College, Kraków, Poland\and
    \small$^{6}$Cracow University of Technology, Faculty of Mechanical Engineering, Kraków, 31-864, Poland\and
    \small$^{7}$AGH University of Krakow, Kraków, Poland.\and \and
	\small$^\ast$Corresponding author. Email: p.moskal@uj.edu.pl
}

\begin{document} 

\maketitle

\begin{abstract} \bfseries \boldmath

Annihilation photons are quantum-entangled in polarization, a phenomenon that has not been exploited in medical diagnostics so far. We present the first in vivo imaging of the degree of quantum entanglement of photons originating from positron–electron annihilation within a human subject. This study utilized the Jagiellonian Positron Emission Tomography (J-PET) scanner, constructed from plastic scintillators. In plastics, annihilation photons interact primarily via the Compton effect, which provides simultaneous information regarding the photon interaction position and time, as well as the photon polarization plane. The patient was injected with a DOTA-TATE radiopharmaceutical labeled with the $^{68}\text{Ga}$ radionuclide. Using the J-PET scanner, we determined the image of the radiopharmaceutical uptake and, simultaneously, the image of the degree of quantum entanglement. The latter was determined from the relative angle between the polarization planes of the annihilation photons. The values of the degree of quantum entanglement extracted for the liver and the spleen are smaller than those predicted for maximally entangled two-photon states, yet larger than expected for separable photons. This demonstration opens new perspectives for the application of quantum entanglement in clinical diagnostics.
\end{abstract}

\section*{Introduction}

State-of-the-art positron emission tomography (PET) is based on the positron-electron annihilation into two photons inside the human body ~\cite{Vandenberghe:2020}. Registering the interaction time and position of these photons within the PET scanner enables the reconstruction of images showing the metabolic rate of molecules to which a positron-emitting radionuclide has been attached~\cite{Alavi:2021}. Standard PET images provide diagnostic information by detailing abnormalities in tissue metabolism~\cite{Schwenck:2023}.

\begin{figure}[!p]
    \centering
    \includegraphics[width=1.0\textwidth]{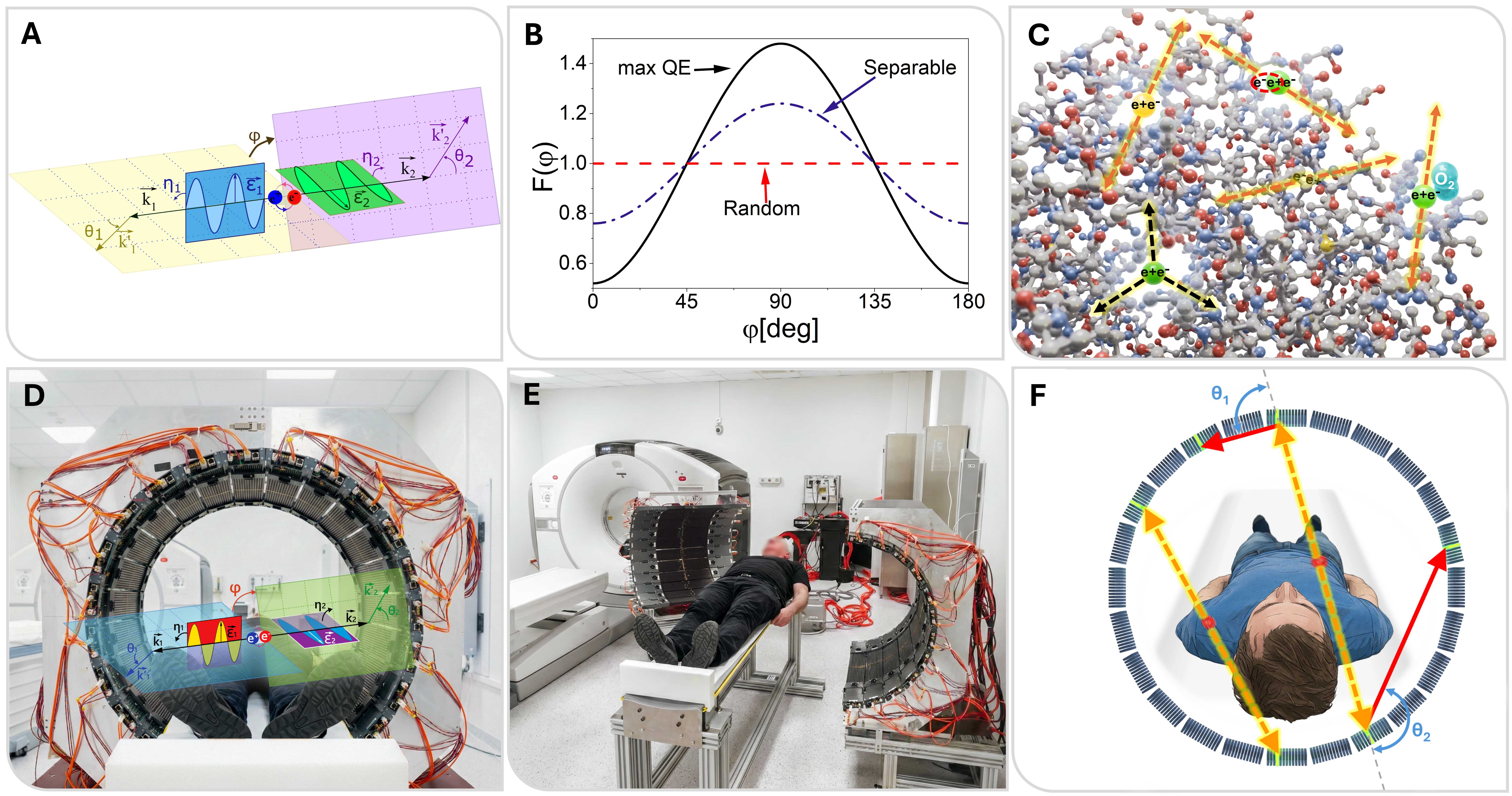}
    \caption{
    {
        \textbf{Principle of quantum entanglement imaging with the J-PET scanner. }
        \textbf{(A)} Schematic of the two-photon annihilation process and subsequent Compton scattering events. The annihilation photon momenta are indicated as $\vec{k}_1$ and $\vec{k}_2$, while the momenta of the scattered photons are denoted by $\vec{k'}_1$ and $\vec{k'}_2$. The polarization vectors ($\vec{\epsilon}_1$, $\vec{\epsilon}_2$), scattering angles ($\theta_1$, $\theta_2$), azimuthal angles ($\eta_1$, $\eta_2$) between the polarization and scattering planes, and the relative angle $\varphi$ between the scattering planes are also indicated.
        \textbf{(B)} Theoretical angular correlation function $F(\varphi)$ for maximally entangled (black solid), separable (blue dashed-dotted), and classically random (red dashed) photon states.
        \textbf{(C)} Molecular-level illustration of positronium (Ps) interactions within a biological environment (hemoglobin molecule). The diagram depicts the annihilation of para-positronium (p-Ps) (yellow circle) into two photons 
        and the annihilation of ortho-positronium (o-Ps) (green circles) into three photons. 
        Also shown are the o-Ps "pick-off" annihilation (red-dashed ellipse), the o-Ps to p-Ps spin conversion mediated by oxygen molecules, and a direct electron-positron annihilation that occurs without the formation of positronium.
        \textbf{(D)} The J-PET tomograph with a patient. A two-photon Compton scattering event is superimposed.
        \textbf{(E)} The modular J-PET scanner shown in its open configuration — split into two halves — facilitating patient access and enabling flexible use alongside conventional CT scanners.
        \textbf{(F)} Trans-axial cross-section of the modular J-PET scanner containing a human phantom, illustrating the events of interest: primary annihilation photons (orange dashed arrows) and the corresponding Compton-scattered photons (red solid arrows) at angles $\theta_1$ and $\theta_2$.
    }  
}
   \label{fig:1}
\end{figure}

Photons from electron-positron annihilation are predicted to be quantum-entangled in polarization~\cite{Bass:2023rmp,Bohm:1957}. The direction of the polarization plane of such high-energy photons (511 keV) can be sensed by determining the plane in which the annihilation photon undergoes Compton scattering off an electron~\cite{KleinNishina:1929}. Therefore, measuring the distribution of the angle between the scattering planes (angle $\varphi$ in Fig. 1A) enables the study of the quantum entanglement of these photons. The distribution of angle $\varphi$ , depends on the degree of quantum entanglement (C$_{QE}$) and the scattering angles ($\theta_1,\theta_2$). In general, the formula describing this distribution reads~\cite{Pryce:1947,Tkachev:2025} (see Methods section):
\begin{equation}
F(\theta_1,\theta_2,\varphi) = 1 - C_{QE} 
\frac{\sin^2\theta_1\ \left(2-\cos{\theta_1}\right)}{2+\left(1-\cos{\theta_1}\right)^3}\cdot \frac{\sin^2\theta_2\ \left(2-\cos{\theta_2}\right)}{2+\left(1-\cos{\theta_2}\right)^3} \cos(2\varphi)
\end{equation}
The annihilation of an electron-positron pair with zero relative angular momentum is predicted to yield maximally entangled photons (C$_{QE}$ = 1), with a $\varphi$ distribution shown as a solid black line in Fig. 1B. For non-entangled separable photons, propagating independently of each other (C$_{QE}$ = 1/2), a weaker correlation is expected, indicated by the dashed-dotted blue line in Fig. 1B.

Recently, two discoveries have been reported that open new perspectives for the possible application of quantum entanglement as a diagnostic indicator~\cite{Moskal:2025SciAdv,Ivashkin:2023}. First, it was demonstrated that in matter, $C_{QE}$ is smaller than predicted for maximally quantum-entangled photons~\cite{Moskal:2025SciAdv}, indicating that it may depend on the annihilation mechanisms (Fig. 1C), and hence on the oxygen concentration and the tissue molecular architecture at the site of annihilation~\cite{Moskal:2026BAMS}.
Second, it was demonstrated that the Compton scattering of an annihilation photon off an electron does not lead to a loss in the degree of quantum entanglement for scattering angles below $\sim~40^\circ$~\cite{Ivashkin:2023,Tkachev:2025,Parashari:2024,Bordes:2024,Caradonna:2024}. In standard PET systems, photons scattered in the body at angles larger than $36^\circ$ are typically discarded; therefore, the scattering of annihilation photons within the body will not dilute the information about the annihilation site environment encoded in the degree of quantum entanglement~\cite{Moskal:2026BAMS}. 

In this work, for the first time we determine the degree of quantum entanglement within the human body. Following a routine PET diagnostic procedure using a $^{68}\text{Ga}$-labeled DOTA-TATE radiopharmaceutical, the patient was examined using a lightweight (60 kg) and portable J-PET scanner constructed from plastic scintillators~\cite{moskal:2024positronium}. Figure 1D displays a photograph of the patient in the scanner, with a superimposed image showing the annihilation photons. The scanner, divided into half-cylinders (Fig. 1E), was rolled up to the patient lying on the diagnostic bed. Annihilation photons interact with plastic scintillators almost exclusively via the Compton effect, making the J-PET scanner well-suited for this study~\cite{Moskal:2018}. Figure 1F schematically shows a cross-section of the scanner with superimposed annihilation (orange-dashed arrows) and Compton-scattered photons (red arrows). Events involving two annihilation photons are recorded for standard PET metabolic imaging to monitor radiopharmaceutical uptake, whereas events including both annihilation and Compton-scattered photons are recorded and used to determine and image the degree of quantum entanglement between photons originating from positron–electron annihilations in the patient.
\section*{Results}
In the reported study, a 48-year-old male patient, administered 145 MBq of [$^{68}$Ga]Ga-DOTA-TATE, was scanned using the J-PET scanner for 20 minutes starting 90 minutes post-injection, following the completion of his routine diagnosis with the GE PET/CT scanner. 

\begin{figure}[!b]
    \centering
    \includegraphics[width=1.0\textwidth]{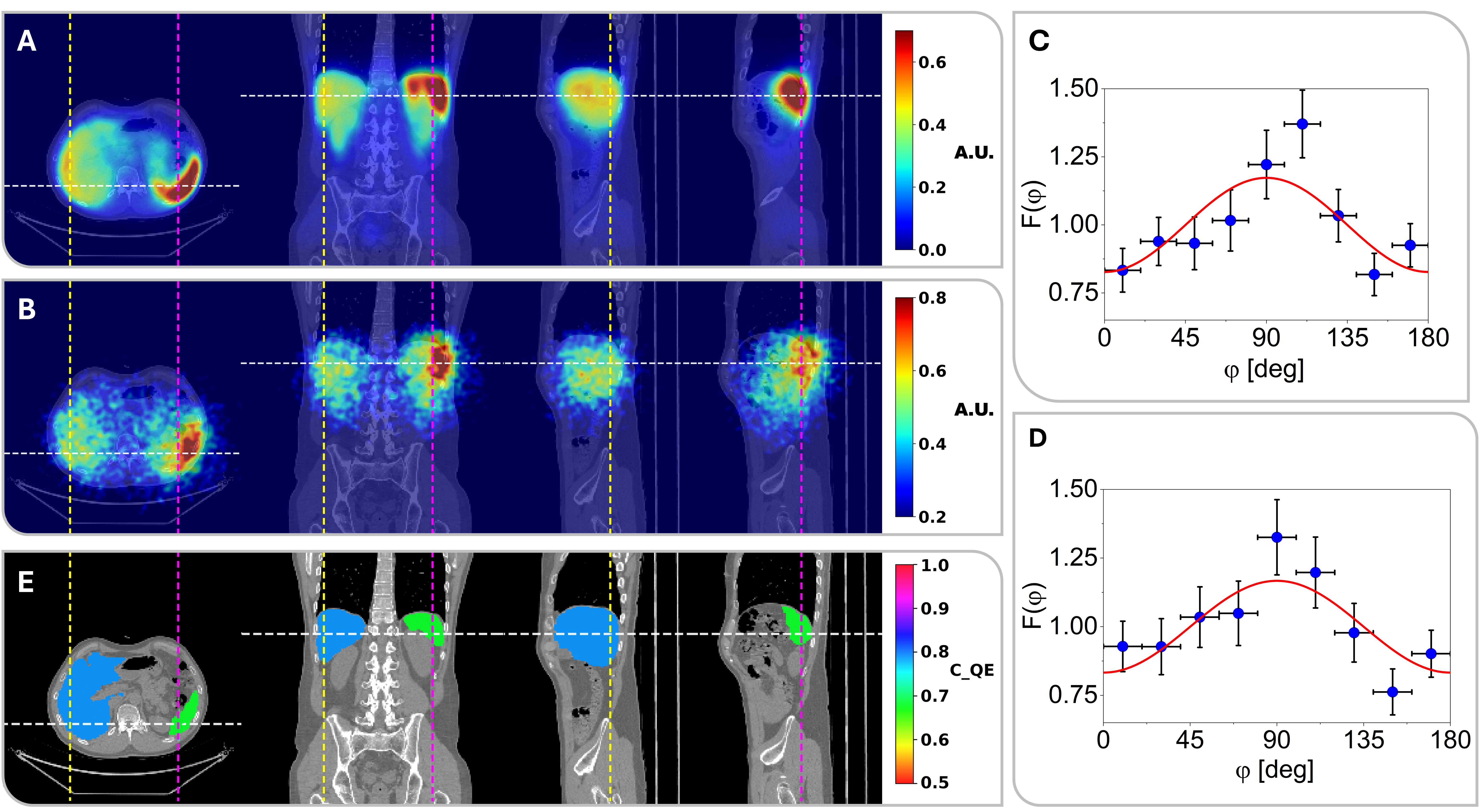}
    
    \caption
    {
       \textbf{Imaging of quantum entanglement in the human body with J-PET.}
       \textbf{(A)} Fused \hspace{5cm} J-PET/CT images showing 2$\gamma$ annihilation activity distribution reconstructed using standard methodologies adapted to J-PET scanner~\cite{Reimund:2026}. J-PET images overlaid on CT images in three anatomical orientations (axial, coronal, and sagittal) are shown. Sagittal cross sections are shown separately for liver and spleen as indicated by yellow and magenta thin dashed lines in axial and coronal planes. 
       \textbf{(B)} Axial, coronal, and sagittal distribution of annihilation points reconstructed from events with  four hit topology selected for quantum entanglement studies.  
       \textbf{(C, D)} Angular correlation function $F(\varphi)$ determined for liver (C) and spleen (D), both centered at scattering angles $\theta_1 = \theta_2 = 110^\circ$ with an angular acceptance of $30^\circ$. Blue points represent experimental data with vertical error bars indicating statistical uncertainties and horizontal bars denoting bin widths. The solid red curves are results of the fit of $F(\varphi)$ function with $C_{QE}$ as a free parameter.   
       \textbf{(E)} The degree of quantum entanglement, $C_{QE}$, superimposed on the CT anatomical images.
    \label{fig:2}
   }
\end{figure}

The main result of the investigation is presented in Fig. 2. The images in the transverse, coronal, and sagittal planes in the upper panel (Fig. 2A) show standard physiological uptake of the administered [$^{68}\text{Ga}$]Ga-DOTA-TATE radiopharmaceutical. The most pronounced uptake is visible in the liver and the spleen. These images were reconstructed on the basis of the registration and identification of $2\gamma$ events, indicated by the orange dashed arrows in Fig. 1F. In addition, events involving the registration of both annihilation and Compton-scattered photons were also identified, indicated in Fig. 1F by the orange dashed arrows and solid-red arrows. These events were selected based on the correlations between the time and position of each of the four interactions within the scanner (details are provided in the Methods section). The interactions of photons within the scintillator are hereafter referred to as 'hits'. Based on the position of each hit, the momentum vectors of the annihilation photons ($\vec{p}_1^{ann}$, $\vec{p}_2^{ann}$) and the scattered photons ($\vec{p}_1^{scat}$, $\vec{p}_2^{scat}$) were calculated and the angle $\varphi$ between the scattering planes was determined for each event. Moreover, based on the time and position of the hits from annihilation photons, the annihilation position ($\vec{r}_{ann}$) was reconstructed. Thus, for each event, a pair consisting of $\varphi$ and $\vec{r}_{ann}$ was determined. Fig. 2B shows the distribution of reconstructed annihilation points for the four-hit events discussed. Although the statistics for these events are limited, the intensity of the annihilation points correlates very well with the intensity (uptake) images shown in Fig. 2A. Based on these images (Fig. 2B), we selected events corresponding to annihilations in the liver and the spleen (see the Methods section). The resulting distributions, corrected for the registration efficiency and random coincidences, are shown in Fig. 2C for the liver and Fig. 2D for the spleen. In both cases, the distributions differ from the predictions for maximally quantum-entangled photons and from those for separable photons. The determined degree of quantum entanglement of photons originating from the annihilations in the liver and spleen of the examined patient is given in Table 1. It amounts to $C_{QE}$ = 0.79 $\pm$ 0.21 for the liver and $C_{QE}$ = 0.76 $\pm$ 0.23 for the spleen. By combining the anatomical images, the $2\gamma$ images (Fig. 2A), and the values of $C_{QE}$, the organ-wise images of the degree of quantum entanglement were determined and are shown in Fig. 2E.

\begin{table}[h]
\centering
    \caption{The degree of quantum entanglement $R_{QE}$ (defined in Methods) and $C_{QE}$ determined for liver and spleen.}
    \vspace{5 mm}
        \begin{tabular}{lcc}
            \hline\hline
            \textbf{Organ} & $R_{QE}$ & $C_{QE}$\\
            \hline
            spleen  & $1.40 \pm 0.14$ & $0.76 \pm 0.23$\\ 
            liver   & $1.42 \pm 0.15$ & $0.79 \pm 0.21$\\
            \hline\hline
        \end{tabular}
    \label{table}
\end{table}

The systematic uncertainties of the $C_{QE}$ determination were estimated to be negligible with respect to the quoted statistical errors. Monte Carlo simulations, described in detail in the Methods section, confirmed high event purity for the liver (97\%) and spleen (94\%) regions determined taking into account the scanner's TOF resolution, organ volumes, and mean uptakes (Table 2). Here purity is defined as the fraction of events from a given organ out of all events identified as originating from this organ.

Systematic variations in the size of the delineated regions by the spatial resolution resulted in $C_{QE}$ values that were consistent within statistical uncertainties. The influence of the random coincidences on the result was also taken into account. The total activity of $^{68}$Ga in a patient at the beginning of the measurement was equal to 57~MBq, resulting in the fraction of randoms of 30.6\% estimated using a delayed time window method~\cite{Manish:2024,Reimund:2026}. This contribution of randoms was taken into account in the analysis. The modification of the $\varphi$ distribution due to the contribution of random coincidences was accounted for in the simulations.
The scatter fraction was estimated to be 32.7\% using a Monte Carlo–based 
scatter correction algorithm developed for the J-PET scanner~\cite{Reimund:2026}. However, it is important to stress that the degree of quantum entanglement, $C_{QE}$, is not affected for annihilation photons that were scattering in the patient at angles lower than about 40$^\circ$~\cite{Moskal:2026BAMS,Tkachev:2025,Caradonna:2024,Parashari:2024,Bordes:2024}, which is the case for the events selected for imaging in this work.

\section*{Discussion and conclusion}
In this work, a measurement of the degree of quantum entanglement of photons from electron–positron annihilation in a human subject is reported for the first time. Ninety minutes after the injection of the [$^{68}\text{Ga}$]Ga-DOTA-TATE radiopharmaceutical, the patient was examined in the modular J-PET scanner (Fig. 1D and 1E). The application of the scanner, constructed from plastic scintillators, enabled the simultaneous determination of (i) the standard PET metabolic images of the patient (Fig. 2A) and (ii) the images (Fig. 2E) of the degree of entanglement extracted from the distribution of the angle between the polarization planes of photons from $e^+e^- \to 2\gamma$ annihilations for the patient's liver and spleen (Fig. 2C and 2D). The values of the degree of quantum entanglement determined, $C_{QE} = 0.79 \pm 0.21$ for the liver and $C_{QE} = 0.76 \pm 0.23$ for the spleen, are consistent with the range of values from 0.784 to 0.890 recently predicted for adipose tissue ($C_{QE} = 0.890$), water ($C_{QE} = 0.867$), and organic liquids such as isopropanol ($C_{QE} = 0.886$), cyclohexane ($C_{QE} = 0.818$), and isooctane ($C_{QE} = 0.784$)~\cite{Moskal:2026BAMS}. Measurements of $C_{QE}$ had not been performed previously for any tissue, not even ex vivo.
Moreover, predictions of $C_{QE}$ do not exist for any tissue other than adipose tissue. Therefore, the values of $C_{QE}$ determined in this work constitute the first report of this type.

The observed non-maximal value of $C_{QE}$ can be interpreted in light of recent findings regarding the non-maximal entanglement of photons from annihilation in a porous-polymer XAD-4~\cite{Moskal:2025SciAdv}. Specifically, this can be explained by assuming that photons created via the pick-off process (Fig. 1C) are separable, whereas photons from other processes remain maximally entangled~\cite{Moskal:2026BAMS}. In the pick-off process, the positron annihilates with an electron from a surrounding molecule, whose spin is uncorrelated with that of the positron. Consequently, photons originating from such a mixed state are not maximally entangled~\cite{Moskal:2026BAMS}. Because the degree of quantum entanglement depends on the fraction of pick-off annihilations, $C_{QE}$ becomes sensitive to tissue structure and biochemical composition (type). This fraction is directly reflected in the mean lifetime of ortho-positronium atoms. Ex vivo studies have shown that this lifetime varies—for instance, between healthy adipose tissue and soft, jelly-like, and friable tumors like cardiac myxoma~\cite{Moskal:2021SciAd,Moskal:2023EJNMMI}—while in vivo studies demonstrate similar variations between healthy brain tissue and glioma tumors~\cite{moskal:2024positronium}. Overall, the mean lifetime of o-Ps in tissue varies between 1.4 and 2.9~ns~\cite{Bass:2023rmp,Moskal:2021SciAd,Moskal:2023EJNMMI,Chen:2012,Jasinska:2017,Zgardzinska:2020,Ahn:2021ACS,Karimi:2023,Avachat:2024,Moyo:2025}. This indicates that the degree of quantum entanglement, $C_{QE}$, may likewise depend on the tissue type~\cite{Moskal:2026BAMS}.

The first determination of the degree of quantum entanglement for the liver and spleen in vivo, presented in this work, opens perspectives for the development of quantum entanglement as a molecular biomarker. Although the measurement statistics for these initial results are low, they can be significantly improved in the near future. For instance, the sensitivity of the total-body J-PET, which is currently under construction, is estimated to be about 3 cps/kBq~\cite{Moskal:2026BAMS} for quantum entanglement imaging. This would be comparable to standard PET imaging with short axial field-of-view systems. In the standard crystal-based PET detectors the double Compton scattering followed by the detection of scattered photons constitutes a few percent of all $2\gamma$ events~\cite{Kozuljevic:2026,Bharathi:2026}. Upgrading highly sensitive total-body PET configurations, currently reaching 174 cps/kBq~\cite{Spencer:2021}, to record these interactions would yield a quantum entanglement imaging sensitivity of roughly 3 cps/kBq.

It is also worth mentioning that (i) recently developed iterative algorithms for positronium lifetime imaging~\cite{Qi:2022,Chen:2023,HuangB:2024,HuangH:2025}, which enable the imaging of parameters independent of the annihilation intensity, can in principle be adapted for imaging the degree of quantum entanglement ($C_{QE}$), and that (ii) there is an intense ongoing development of new generation PET detectors in view of its application for simultaneous registration of annihilation and Compton scattered photons~\cite{Yoshihara:2017,Moskal:2018IEEE,Makek:2020,Shimazoe:2020,Yoshida:2020,Tashima:2022,Kim:2023,Romanchek:2023,Romanchek:2024,Kim:2024,Bordes:2024,Kozuljevic:2026,Bharathi:2026,Schimazoe2026review}.

In summary, the angular distributions between the scattering planes of annihilation photons in the patient's liver and spleen were determined. These distributions were used to extract the degree of quantum entanglement for these organs. The obtained degree of quantum entanglement is larger than expected for separable photons and lower than predicted for maximally entangled photons. These results demonstrate for the first time the feasibility of measuring and imaging the degree of quantum entanglement in annihilation photons originating from patients during positron emission tomography. This finding motivates the further development of detectors for the construction of PET systems capable of simultaneously capturing annihilation and Compton-scattered photons, and encourages future studies aimed at uncovering medically useful information gained by imaging the degree of quantum entanglement.

We envision that quantum entanglement imaging has the potential to transcend the current PET diagnostic paradigm, opening a promising path toward developing new biomarkers for assessing the molecular environment of tissues in vivo.

\noindent

\section*{Methods}

\subsection*{\textbf{Description of the ethical committee, patient and radiopharmaceutical}}

The study was performed at the Chair and Department of Endocrinology of the Jagiellonian University Medical College and University Hospital in Kraków. The study was carried out in March–April 2024 and was approved by the ethical committee under agreement No. 1072.6120.92.2023 (NCT06242119). The participant was fully informed about the study and provided written informed consent for participation in the study.

The patient was a 48-year-old male diagnosed with a well-differentiated pancreatic neuro-endocrine tumor (NET G1, Ki‑67 <3\%). In October 2022, he underwent segmental pancreatic resection, with histopathology confirming pT1 N0 R0 disease according to AJCC 2017 criteria. To evaluate the postoperative status and exclude residual or metastatic disease, he was injected with 145.42 MBq of [$^{68}$Ga]Ga-DOTA-TATE. 55 minutes post-injection, the patient was scanned on a commercial GE Discovery MI Gen 2 PET/CT scanner for a duration of 28 minutes. After completion of the PET/CT acquisition, the patient was asked to void bladder activity, and 90 minutes post-injection, the patient was scanned on the J-PET tomograph for 20 minutes (abdominal cavity).

\section*{Formalism on the degree of quantum entanglement}

The distribution of the relative angle $\varphi$ between the scattering planes (Fig. 1A) depends on the  scattering angles $\theta_1$ and $\theta_2$, and may be expressed by the following formula:
\begin{equation}
F(\theta_1,\theta_2,\varphi) = 1 - V(\theta_1,\theta_2) \cos(2\varphi), 
\end{equation}
where V is referred to as the visibility describes the interference contrast:
\begin{equation}
V(\theta_1, \theta_2) = \frac{F(\theta_1, \theta_2, \varphi = 90^\circ) - F(\theta_1, \theta_2, \varphi = 0^\circ)}{F(\theta_1, \theta_2, \varphi = 90^\circ) + F(\theta_1, \theta_2, \varphi = 0^\circ)}.
\end{equation}
Visibility V includes the dependence on the scattering angles ($\theta_1,\theta_2$) and the degree of the quantum entanglement $C_{QE}$ that can be factorized~\cite{Tkachev:2025}:
\begin{equation}
V = C_{QE} \cdot A(\theta_1) A(\theta_2), 
\end{equation}
where $A(\theta_1)$ and $A(\theta_2)$ represent the analyzing powers of a Compton polarimeter.
For 511 keV photons originating from $e^+e^-\rightarrow \gamma\gamma$ annihilation, the angular dependence $A(\theta)$ is given by:
\begin{equation}
A(\theta) = \frac{\sin^2\theta\ \left(2-\cos{\theta}\right)}{2+\left(1-\cos{\theta}\right)^3} 
\end{equation}
This dependence of the $\varphi$ angle distribution on the scattering angles $\theta_1$ and $\theta_2$ implies that the difference between quantum entangled and separable states is at most prominent when the scattering angles are $\theta_1 = \theta_2 = 82^{\circ}$~\cite{Pryce:1947}.

The degree of quantum entanglement expressed in the value of $C_{QE}$ is independent of the scattering angles $\theta_1$ and $\theta_2$ \cite{Tkachev:2025}. For maximally quantum entangled photons $C_{QE}$ = 1 and for separable photons, propagating independently, $C_{QE}$ = 1/2. $C_{QE}$ is a newly introduced measure of entanglement~\cite{Tkachev:2025,Moskal:2026BAMS}. Therefore, in this manuscript, we also use parameter $R_{QE}$, which was used  as a measure of the degree of entanglement in the prior research~\cite{Moskal:2018,Ivashkin:2023,Parashari:2024,Bordes:2024,Romanchek:2024,Moskal:2025SciAdv}.  $R_{QE}$ is defined as the ratio of the probabilities of scattering at $\varphi = 90^\circ$ and $\varphi = 0^\circ$~\cite{Sharma:2022Acta}:
\begin{equation}
R_{QE}(\theta_1, \theta_2, C_{QE}) = \frac{1 + C_{QE} A(\theta_1) A(\theta_2)}{1 - C_{QE} A(\theta_1) A(\theta_2)}
\end{equation}
The events used in this study involve scattering angles around $\theta_1 = \theta_2 = 110^\circ$. At these angles, the ratio $R_{QE}$ is equal to  1.56 for maximally quantum entangled photons, whereas for separable photons, $R_{QE} = \text{1.24}$.
The degree of quantum entanglement $C_{QE}$ is related to $R_{QE}$ by the following formula:

\begin{equation}
C_{QE} = \frac{R_{QE} - 1}{A(\theta_1) A(\theta_2) (R_{QE} + 1)}
\end{equation}

\section*{The Modular J-PET Scanner}

Measurements were carried out using the modular J-PET Tomography scanner (Figs. 1D-1E), a portable PET system constructed from plastic scintillator strips~\cite{moskal:2024positronium}.
The scanner comprises 24 independent detection modules arranged in a cylindrical geometry with a bore diameter of 73.9\,cm and a 50\,cm axial field of view. Each module consisting of 13 plastic scintillator strips of dimensions $6\,\text{mm} \times 24\,\text{mm} \times 500\,\text{mm}$, read out at both ends by $1\times4$ silicon photomultiplier (SiPM) arrays~\cite{Tayefi:2024BAMS,Kacprzak:2026BAMS}. Unlike conventional crystal-based PET detectors, where 511\,keV annihilation photons interact primarily via the photoelectric effect, 511\,keV photons in plastic scintillators interact predominantly via Compton scattering~\cite{Moskal:2025SciAdv}, with photoelectric effect at the negligible level of 6.3 $\times$ 10$^{-5}$~\cite{Moskal:2021pmb}.
This is a key distinction: in each Compton interaction, the photon scatters at an angle correlated with its linear polarization direction. 
In the modular J-PET scanner used in this study, at least one of the annihilation photons undergoes a second scattering in about 33\% of cases, while both annihilation photons scatter twice in 4.5\% of cases, enabling the determination of the $\varphi$ angle. Thus, J-PET detector registers not only the direction of propagation of the annihilation photons but also information about their polarization -- a capability that is absent in conventional crystal-based PET systems and that is essential for the present study~\cite{Moskal:2025SciAdv, Moskal:2024NatComm}. The two-dimensional $(x,y)$ hit position is determined from the known scintillator strip location, while the axial ($z$) coordinate is extracted from the time difference of signals arriving at opposite ends of the strip~\cite{Moskal:2014nim,moskal:2024positronium}. The analog signals are processed directly by the field-programmable gate array (FPGA) by recording signal crossings at two constant thresholds to provide a time-over-threshold (TOT), a measure of the energy deposition~\cite{Sharma:2020EJNMMI}. Data acquisition operates in a triggerless and reconfigurable mode, recording all coincident hits within a programmable time window ($50~\mu s$) without hardware-level event filtering~\cite{Korcyl:2014,Korcyl:2018}. This architecture enables simultaneous standard PET imaging~\cite{Manish:2024,Reimund:2026}, positronium lifetime imaging~\cite{Moskal:2018IEEE,Moskal:2021SciAd,moskal:2024positronium}, and measurement of the polarization correlations of annihilation photon pairs via Compton scattering within the detector volume itself~\cite{Moskal:2018,Moskal:2025SciAdv}.

\section*{Data Analysis}

\subsection*{Event Reconstruction and Hit Classification}

Data were collected continuously for 20\,min and analyzed using the dedicated J-PET analysis framework~\cite{moskal:2024positronium} built upon the ROOT data analysis toolkit~\cite{ROOT:1997}. 
Data acquisition in triggerless mode enabled the simultaneous registration of primary and secondary scattering of annihilation photons. Event selection was based on the position, time, and energy deposition of the annihilation and scattered photons. Two classes of events were selected (Fig. 1F): (i) events corresponding to registration of two annihilation photons, and (ii) events with registration of two annihilation photons and additional Compton scattered photons. Events useful for the study of polarization correlations (indicated in Fig. 1F) comprise at least four hits registered within a 5\,ns time window: two hits from primary annihilation photons (511\,keV) and two hits from their corresponding Compton-scattered photons, which carry lower energy (e.g., 
218\,keV for scattering at 110 degrees). 
The  distribution of multiplicity of hits in the events is shown in Fig.~3A.
For the second class of events, used to determine the $\varphi$ angle, only events containing at least four hits were retained for further analysis as indicated in blue in Fig.~3A.
Each hit was characterized by its reconstructed position along the scintillator strip, its registration time, and its time-over-threshold (TOT) value, which is correlated with the energy deposited in the scintillator~\cite{Sharma:2020EJNMMI}.
The axial position $z$ was restricted to $|z| < 23\,\text{cm}$. Hits were then classified by TOT into two categories (Fig. 3B). Annihilation photon candidates were required to satisfy $5.0\,\text{ns~V} < \text{TOT} < 7.5\,\text{ns~V}$, where the upper limit corresponds to the Compton edge of 340\,keV for 511-keV photons~\cite{Moskal:2025SciAdv}. Scatter photon candidates were required to satisfy $2.0\,\text{ns~V} < \text{TOT} < 5.0\,\text{ns~V}$, consistent with the lower energy depositions expected from Compton-scattered photons. 
Events for the determination of $\varphi$ angle were required to contain at least two annihilation photon candidates and exactly two scatter photon candidates.
\begin{figure}
    \centering
    \includegraphics[width=1.0\textwidth]{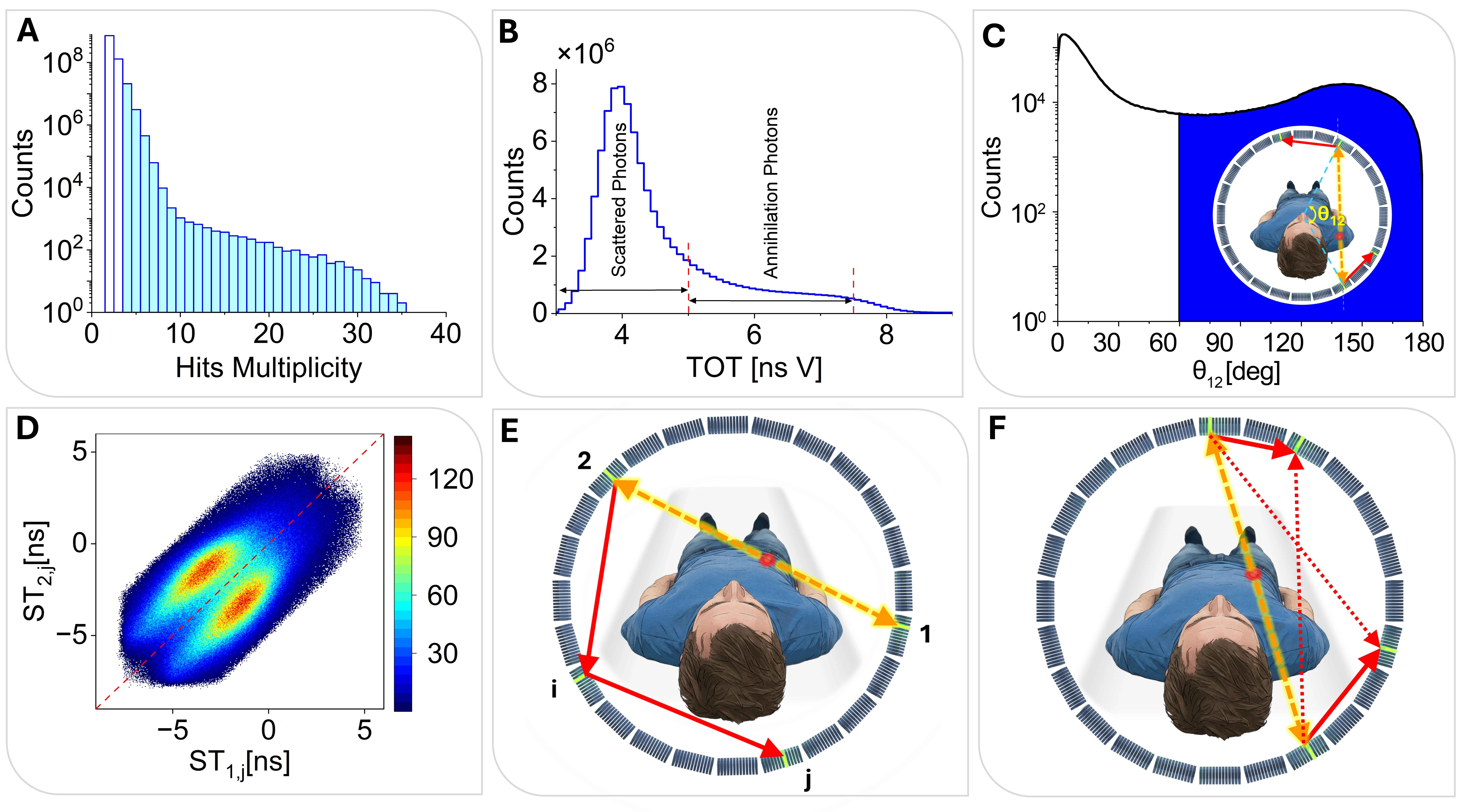}
    \caption
    {
        \textbf{Event selection criteria and examples of background events.}
        \textbf{(A)} Distribution of hit multiplicity per event. Events with multiplicity $< 2$ are rejected at the pre-selection stage. 
         \textbf{(B)} Distribution of time-over-threshold (TOT) used to select scatter and annihilation photons.
         \textbf{(C)} Distribution of the relative angle $\theta_{12}$ between vectors indicating interaction position of annihilation photons, calculated with respect to the scanner center. The distribution from the first selection stage before selection described in the second stage is shown.
        \textbf{(D)} Two-dimensional distribution of scatter test variables $ST_{2,j}$ versus $ST_{1,j}$, used to assign each $j$-th scattered photon to either the first or second primary annihilation photon. The dashed red diagonal defines the assignment boundary: events below are attributed to the 1\textsuperscript{st} annihilation photon and those above to the 2\textsuperscript{nd}. 
        \textbf{(E,F)}
        Cross-section of the modular J-PET tomograph with the human phantom, illustrating a representative background event in which a single annihilation photon (dashed orange arrow 2) undergoes double Compton scattering, indicated by red arrows $i$ and $j$ (E), and a background event due to the misassignment of scattered photons, where the dotted arrows were mixed with the red arrows (F).
    }   
    \label{fig:3}
    
\end{figure}
\subsection*{Primary Event Selection and Scatter Photon Assignment}

From the hits fulfilling conditions for annihilation candidates, all possible pairs were formed and subjected to two sequential kinematic criteria. The time difference between the two annihilation hits was required to satisfy $\Delta t < 2.5\,\text{ns}$, and the opening angle between the position vectors of the two hits with respect to the detector centre was required to exceed $60^\circ$ (Fig. 3C). Only events containing exactly one such surviving pair were selected for further analysis.

For each selected annihilation pair $(A_1,\,A_2)$, the two scatter candidates were assigned to their respective annihilation photons using a scatter test (Fig.~3D), defined as:
\begin{equation}
    \mathrm{ST}_{k,j}
    = \left(t_j - t_k\right)
      - \frac{|\vec{r}_j - \vec{r}_k|}{c},
    \label{eq:scatter_test}
\end{equation}
where $k \in \{1,2\}$ denotes the annihilation photon, $j$ denotes the scatter candidate, $t$ and $\vec{r}$ denote the hit time and position, respectively, and $c$ is the speed of light. For an ideal detector, $\mathrm{ST}_{k,j} = 0$ if the $j$-th scatter originates from the $k$-th annihilation photon. Fig.~3D shows the scatter test plot where two well distinct groups of events are well visible. The upper one corresponds to the case where the scatter originates from the scattering of $A_2$ and the lower from scattering of $A_1$. A scatter candidate $j$ was assigned to $A_1$ if $\mathrm{ST}_{1,j} > \mathrm{ST}_{2,j}$, and to $A_2$ otherwise (red line in Fig.~3D).
Events in which either annihilation photon had no scatter candidate assigned were rejected.

\subsection*{Second stage selection for background suppression}

A set of criteria was applied to suppress accidental coincidences, geometrically degenerate configurations, and misassigned scatter events. Examples of such events are shown in Figs. 3E and 3F. Fig. 3E illustrates an event where one annihilation photon underwent two secondary scatterings, while the other annihilation photon did not scatter a second time. Fig. 3F shows an example of a wrong assignment of the scatter photons, which causes an incorrect estimation of the scattering angles. These events can be suppressed by requiring that scattered hits are not close to each other. Specifically, the annihilation-to-scatter distance (ASD) for each assigned photon-scatter pair was required to exceed $10\,\text{cm}$. This ensures that the scatter hit is spatially separated from the annihilation hit, thereby rejecting cases where both interactions occur in the same or adjacent scintillator strips. Furthermore, the scatter-to-scatter distance (SSD) was required to exceed $30\,\text{cm}$ to ensure sufficient spatial separation between the two reconstructed Compton scattering vertices and to suppress cases where a single photon could mimic both scatter hits (Fig. 3E).

\subsection*{Annihilation Point Reconstruction and Region of Interest Selection}

The annihilation point for each selected event was reconstructed from the time difference between the two annihilation hits along the line of response (LOR) \cite{moskal:2024positronium}. The reconstructed annihilation points from all events fulfilling the selection criteria  were accumulated into a three-dimensional spatial distribution and overlaid on the patient CT scan (Fig.~2B). 
The spleen and the liver were identified as two spatially distinct regions of elevated annihilation density.

To isolate events originating from each region independently, a three-dimensional ellipsoidal region of interest (ROI) was defined around each identified cluster. An event was accepted if its reconstructed annihilation point $\vec{r} = (x,\,y,\,z)$ satisfied
\begin{equation}
    \left(\frac{x - x_0}{a}\right)^2
    + \left(\frac{y - y_0}{b}\right)^2
    + \left(\frac{z - z_0}{c}\right)^2 < 1.
    \label{eq:ellipsoid}
\end{equation}
For the spleen, the ellipsoid was centered at $(x_0,\,y_0,\,z_0) = (11.5,\,-7.0,\,-7.50)\,\text{cm}$ with semi-axes $(a,\,b,\,c) = (5.0,\,6.0,\,9.0)\,\text{cm}$. For the liver, the center was at $(-5.0,\,-4.0,\,-9.0)\,\text{cm}$ with semi-axes $(6.0,\,9.0,\,7.0)\,\text{cm}$. The ellipsoid boundaries were determined by visual inspection of the reconstructed annihilation point distribution overlaid on the CT anatomy (Fig.~4).

\begin{figure}[h]
    \centering
    \includegraphics[width=.9\textwidth]{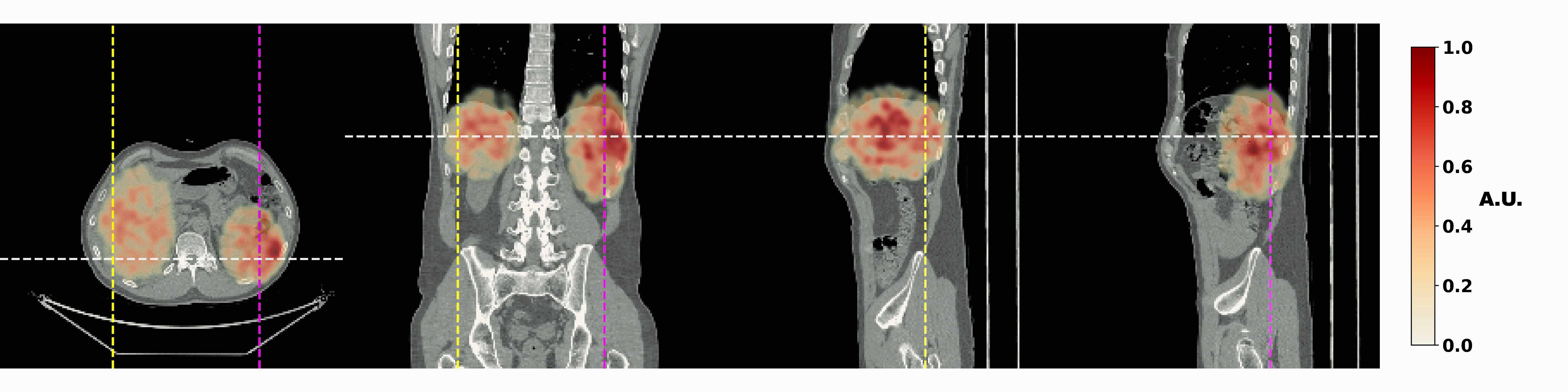}
    \caption
    {
          \textbf{Spatial distribution of events selected for determining the degree of quantum entanglement for liver and spleen.} Axial, coronal, and sagittal planes showing annihilation point distributions 
           overlaid on CT images. These distributions correspond to the regions used to select events for annihilations in liver and spleen. Sagittal cross sections are shown separately for liver and spleen as indicated by yellow and magenta thin dashed lines in axial and coronal planes. 
      \label{fig:4}
    }
\end{figure}
\subsection*{Polarization Correlation}

For each accepted event, the normal to the Compton scattering plane of each annihilation photon was computed as
\begin{equation}
    \hat{e}_i = \hat{k}_i \times \hat{k}_i',
    \label{eq:plane_normal}
\end{equation}
where $i \in \{1,2\}$ labels the two annihilation photons, and $\hat{k}_i$ and $\hat{k}_i'$ denote the directions of the incoming and outgoing photon at the Compton interaction, respectively. The angle $\varphi$ between the two scattering planes was
then determined as the angle between $\hat{e}_1$ and $\hat{e}_2$.

To maximize the sensitivity to photon polarization, only events in which both Compton scattering angles are in the range where the analyzing power of the Compton polarimeter is significant were retained. A circular cut was applied in the joint
$(\theta_1,\,\theta_2)$ space:
\begin{equation}
    \sqrt{(\theta_1 - 110^\circ)^2 + (\theta_2 - 110^\circ)^2}
    \leq 30^\circ.
    \label{eq:theta_cut}
\end{equation}

The resulting efficiency-corrected $\varphi$ distributions for the spleen and liver regions are shown in Fig.~2C and Fig.~2D, respectively, together with the fitted function $F(\varphi) = A(1 - B\cos(2\varphi))$.

\subsection*{Monte Carlo Simulation and Efficiency Correction}

The corrections for efficiency and geometrical acceptance of the detection system were determined by means of the Monte-Carlo based simulations taking into account full scanner geometry and material composition as described in detail and validated in the previous works~\cite{Moskal:2018,Moskal:2025SciAdv}. To model the distribution of entangled photons, we adapted the simulated dataset from the separable state, applying a preselection filter to isolate events that match the expected theoretical distributions of entanglement~\cite{Moskal:2018,Moskal:2025SciAdv}.

Dedicated Monte Carlo simulations were performed separately for each activity region using Geant4~\cite{Geant4:2016}, with the full modular J-PET detector geometry implemented. For the spleen and liver, the positron source was modeled as a water-filled ellipsoidal volume.

In both cases the simulation volume was chosen to generously encompass the corresponding analysis ROI, ensuring that edge effects in the efficiency estimation are avoided. Simulated events were processed through an identical
reconstruction and selection pipeline as the data, including all hit classification, kinematic, quality, spatial, and scattering angle cuts, yielding reconstructed and true-level Monte Carlo samples for each region separately. The simulations accounted also for the random coincidences by taking into account the activity of the patient (57~MBq) at the beginning of the examination in the J-PET scanner.

The raw $\varphi$ distribution is distorted by the non-uniform detector acceptance and reconstruction efficiency as a function of angle $\varphi$ and scattering angles $\theta_1$ and $\theta_2$. To correct for this, the efficiency was estimated using two methods. In the first, the efficiency $\varepsilon_{1D}(\varphi)$ was treated as a function of $\varphi$ alone; in the second, $\varepsilon_{3D}(\theta_1,\,\theta_2,\,\varphi)$ was evaluated
as a function of the joint variables $ (\theta_1,\,\theta_2,\,\varphi)$:
\begin{align}
    \varepsilon_{1D}(\varphi)
    &= \frac{N_\mathrm{reco}(\varphi)}
            {N_\mathrm{true}(\varphi)},
    \label{eq:1Defficiency} \\[4pt]
    \varepsilon_{3D}(\theta_1,\,\theta_2,\,\varphi)
    &= \frac{N_\mathrm{reco}(\theta_1,\,\theta_2,\,\varphi)}
            {N_\mathrm{true}(\theta_1,\,\theta_2,\,\varphi)},
    \label{eq:3Defficiency}
\end{align}
where $N_\mathrm{reco}$ and $N_\mathrm{true}$ are the number of reconstructed and true-level simulated events in each bin, respectively. In both cases, each data event was weighted by the inverse of the efficiency evaluated at its reconstructed angles. For the one-dimensional correction the weight depends only on the reconstructed $\varphi$, so the corrected spectrum follows directly from the weighted events in $\varphi$ histogram. For the three-dimensional correction the weight depends on the full combination $(\theta_1,\,\theta_2,\,\varphi)$, and the corrected $\varphi$ spectrum is obtained by projecting the three-dimensionally weighted distribution onto the $\varphi$ axis. This three-dimensional correction ensures that each event is corrected for
the efficiency appropriate to its specific combination of angles $(\theta_1,\,\theta_2)$, rather than an average efficiency, thereby avoiding any bias introduced by correlations between the detector response and the
measured $\varphi$ distribution. The $\varphi$ spectra
obtained with the two methods agree within statistical uncertainties, confirming that such correlations are small and that the simpler one-dimensional correction is sufficient to recover the underlying $\varphi$ distribution.

The efficiency-corrected $\varphi$ distributions were fitted with the
function:
\begin{equation}
    F(\varphi) = A\left(1 - B\cos(2\varphi)\right).
    \label{eq:fit_func}
\end{equation}
The free parameters $A$ and $B$ were estimated through a non-linear least-squares minimization employing the Levenberg-Marquardt method, with weights derived from propagated per-bin uncertainties. The parameter $B$
describes the amplitude of the cosine distribution, and the
degree of entanglement $R_{QE}$ was calculated as:
\begin{equation}
    R_{QE} = \frac{F(90^\circ)}{F(0^\circ)} = \frac{1 + B}{1 - B},
    \label{eq:R}
\end{equation}
and then $C_{QE}$ was determined using equation~(7).

\subsection*{\textbf{Description of the PET images obtained with the modular J-PET scanner}}

Following the selection of the two annihilation photons, described in the above subsections, the interaction position and timing information were exported in list-mode format for image reconstruction using the CASToR package~\cite{Merlin:2018}.

Image reconstruction (Fig. 2A) was performed using a time-of-flight list-mode MLEM (TOF LM-MLEM) algorithm with 10 iterations. The TOF uncertainty was modelled using a Gaussian kernel with an FWHM of 490~ps~\cite{Kacprzak:2026BAMS}, and a multi-Siddon projector with 10 rays was employed. Images were reconstructed on a $200 \times 200 \times 200$ voxel grid with a voxel size of $2.5 \times 2.5 \times 2.5$~mm$^3$. Post-reconstruction, a three-dimensional Gaussian filter with an FWHM of 5~mm was applied. Additionally, corrections were applied during reconstruction, including sensitivity, attenuation, random, and scatter corrections. CT data was used for the attenuation and scatter correction. This data was obtained from the PET/CT examinations performed using a GE Discovery MI Gen 2 PET/CT. Further details on the implementation of these correction techniques for the J-PET can be found in~\cite{Manish:2024, Reimund:2026}. 

\section*{Systematic uncertainties}

To estimate systematic uncertainty, we varied the size of the delineated region used to select events for the spleen and liver, extending and decreasing the range in each dimension by spatial resolution of the determination of the annihilation point ($\text{FWHM} \approx 7.4\text{ mm}$). The resulting changes in the values of $C_{QE}$ and $R_{QE}$ were less than the statistical uncertainties of their determination.

\begin{table}[h!]
\centering
\caption{Volume and standard uptake values (SUV) determined for the examined patient based on the PET/CT images.}
\begin{tabular}{lccc}
\hline
Organ & Volume (mL) & SUVmean & SUVmax \\
\hline
Spleen & 195.26 & 20.13 & 32.25 \\
Right Kidney & 136.71 & 7.62 & 17.02 \\
Left Kidney & 157.45 & 7.67 & 14.85 \\
Liver & 1279.84 & 7.12 & 22.36 \\
\hline
\end{tabular}
\label{tab:organ_measurements}
\end{table}

The purity of the sample from a given region of interest was also estimated. The Monte Carlo simulations were used to estimate that within the regions delineated to determine the $\varphi$ distributions, 97\% of the events contributing to Fig. 2C originated from the liver (with 3\% from the kidney), while 94\% of the events contributing to Fig. 2D originated from the spleen (with 6\% from the kidney). This estimation took into account the 490 ps TOF resolution of the J-PET scanner~\cite{Kacprzak:2026BAMS}, as well as the volumes and mean uptakes of the liver, spleen, and kidneys (Table 2), following the detailed procedure described in reference~\cite{ManishPhD:2026}. 

We have also performed correction for efficiency calculating it as a function of $\varphi$ only, instead of using the three dimensional correction defined in equation (\ref{eq:3Defficiency}). The obtained values of $C_{QE}$ differ by less than the statistical uncertainty.

\newpage


\clearpage 


\section*{Acknowledgments}
\paragraph*{Funding:}
We acknowledge support from the National Science Centre of Poland through grants 
MAESTRO no. 2021/42/A/ST2/00423 (P.M.);
OPUS no. 2021/43/B/ST2/02150 (P.M.); 
OPUS24+LAP no. 2022/47/I/NZ7/03112 (E.Ł.S.);
SONATA no. 2023/50/E/ST2/00574 (S.S.);
PRELUDIUM no. 2024/53/N/ST2/04279 (D.K.).
The Ministry of Science and Higher Education through grant no. IAL/SP/596235/2023 (P.M.) and  SPUB/SP/627733/2025 (E.Ł.S.),
European Union within the Horizon Europe Framework Programme through ERC Advanced Grant POSITRONIUM no. 101199807 (P.M.),
The SciMat and qLife Priority Research Areas budget under the program Excellence Initiative – Research University at Jagiellonian University (P.M. and E.Ł.S.).
We also acknowledge Polish high-performance computing infrastructure PLGrid (HPC Center: ACK Cyfronet AGH) for providing computer facilities and support within computational grant no. PLG/2024/017688 and PLG/2025/018762 (M.S.).
\paragraph*{Author contributions:}
The experiment was conducted using the Jagiellonian Positron Emission Tomograph (J-PET). The J-PET scanner, the techniques of the experiment and this study were conceived by P.M. The performed medical experiment was planned by P.M. and E.Ł.S. The ethical committee consent was secured by A.H.D, M. O., A.S.S and E.Ł.S.  The preparation of pharmaceuticals, planning of patient examination and imaging protocols, patient preparation, imaging, and therapy management were done by A.H.D, M. O., and A.S.S. The data analysis was conducted by D.K. Signal selection criteria were developed by P.M. and D. K., and verified by P. M. and S.S. PET images reconstruction and corrections were performed by M.D. Authors: P.M., D.K., S.S, E.Y.B., A.B., N.C., J.C., C.C., E.C., M.D., A.D., A.H.D., J.H., S.J., K. Kacprzak, T. Kaplanoglu., Ł.K., K.Kasperska, G.K., T. Kozik, A.K.V., B.L., E.L., F.L., J.M.S., S.M., W.M., S.N., A.P., B.R., M.S., A.S.S., A.S., T.S., S.T., P.T., K.T.A., K.V.E., and E.Ł.S., participated in the construction, commissioning, and operation of the experimental setup in hospital.
S.N. and G. K. optimized the working parameters of the detector. K. Kacprzak performed timing and energy calibration of the detector. E.C. developed and operated short- and long-term data archiving systems and the computer center of J-PET. S.S. established relation between energy loss and TOT and dependence of detection efficiency on energy deposition.  
P.M. and E.Ł.S. managed the whole project and secured the main financing.  The results were interpreted by P.M. The manuscript was prepared by P.M. with the help of D.K. and was then edited and approved by all authors.

\paragraph*{Competing interests:}
P. Moskal is a holder of patents on PET from plastic scintillators and on positronium imaging.
Other authors have no competing interest to declare.
\paragraph*{Data and materials availability:}
All data needed to evaluate the conclusions in the paper are present in the paper and/or the Supplementary Materials.

\end{document}